\documentclass[apl,twocolumn,showpacs]{revtex4}
\usepackage{graphicx} 
\newcommand{\etal}{{ \it et al. }}

\usepackage{amsmath,amsthm,amssymb,mathrsfs}
\usepackage{bm}
\usepackage{subfigure}

\newcommand{\D}{{\rm d}}
\newcommand{\E}{{\rm e}}
\newcommand{\I}{{\rm i}}
\renewcommand{\Re}{{\rm Re }}
\renewcommand{\Im}{{\rm Im }}

\begin{document}

\title{Critical current of spin transfer torque-driven magnetization
  dynamics in magnetic multilayers}

\author{Tomohiro Taniguchi$^{1,2}$ and Hiroshi Imamura$^{1}$\footnote{\vspace{-3ex}Corresponding author. Email address: h-imamura@aist.go.jp}}
 \affiliation{
 ${}^{1}$ 
 Nanotechnology Research Institute, AIST, Tsukuba, Ibaraki 305-8568, Japan, \\ 
 ${}^{2}$ 
 Institute of Applied Physics, University of Tsukuba, Tsukuba, Ibaraki 305-8573, Japan
 }

\date{\today} 
\begin{abstract}
 {
   The critical current of the spin transfer torque-driven magnetization
   dynamics was studied by taking into account both spin pumping 
   and the finite penetration depth of the transverse spin current. 
   We successfully reproduced the recent experimental results obtained
   by Chen et al. [Phys. Rev. B {\bf 74}, 144408 (2006)] and 
   found that the critical current remains finite even
   in the zero-thickness limit of the free layer.
   We showed that the remaining value of the critical
   current is determined mainly by spin pumping. 
   We also showed that we could control the critical current by varying
   the spin diffusion length of the nonmagnetic electrode adjacent to the
   free layer.
 }
\end{abstract}

\maketitle


Spin transfer torque (STT) is the torque due to the transfer of
transverse 
spin angular momentum from
the conducting electrons to the magnetization of a ferromagnet
\cite{slonczewski96,berger96}.  
The STT in magnetic multilayers
such as the current perpendicular-to-the-plane giant magneto-resistive 
(CPP-GMR) \cite{katine00,carva06,haney07}
and tunnel magneto-reisistive (TMR) \cite{heiliger08,kubota08,sankey08} spin valves 
has been investigated intensively 
because STT-driven magnetization dynamics is 
a promising technique to operate the spin-electronics
devices such as magnetic random access memories and microwave oscillators.  
One of the main obstacles in developing 
STT-based spin-electronics
devices is the high critical current density.
The critical current density required to induce the STT driven magnetization
dynamics in CPP-GMR spin valves is as high as $10^{6} \! - \! 10^{8}$ A/cm${}^{2}$ 
\cite{kiselev03,seki06,deac06,chen06}.

 On the other hand, the CPP-GMR spin-valve is one of the promising
 candidates for the read head for ultra-high-density magnetic
 recording \cite{tanaka02,takagishi02}. 
 It is known that STT-driven magnetization dynamics produces noise, and that 
 low critical current density is required for the read head
 application\cite{zhu04}.  Therefore, it is natural to ask how to control the 
 critical current density of STT-driven magnetization dynamics in
 CPP-GMR spin valves.


STT was first proposed by Slonczewski
\cite{slonczewski96} and independently by Berger \cite{berger96} in 1996. 
In Slonczewski's theory the critical current of STT-driven
magnetization dynamics is expressed as \cite{grollier03,morise05},
\begin{equation}
  I_{\rm c}
  =
  \frac{2eMSd}{\hbar\gamma\eta}
  \alpha_{0}\omega ,
  \label{eq:critical_current1}
\end{equation}
where $e$ is the elementary charge and 
$\hbar$ is the Dirac constant. 
$M$, $S$, $d$, $\gamma$ and $\alpha_{0}$ are 
the magnetization, cross-section area, 
thickness, gyromagnetic ratio and 
intrinsic Gilbert damping constant of the free layer, respectively. 
$\omega$ is the angular frequency of the magnetization around the
equilibrium point.  The transverse spin polarization coefficient
$\eta$ depends only on the relative angle of the magnetizations of 
the fixed and free layers \cite{slonczewski96,grollier03}.
According to Slonczewski's theory, we can control the
critical current by varying the thickness of the free layer, $d$ and 
the critical current vanishes in the limit of $d\to 0$.


\begin{figure}
  \centerline{\includegraphics[width=0.95\columnwidth]{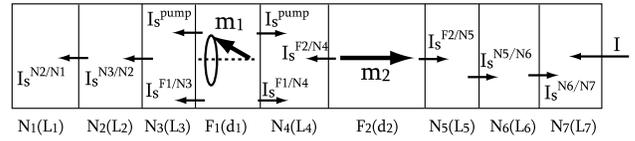}}
  \caption{
    The nonmagnetic(N) / ferromagnetic(F) multilayer, {\it i.e.},
    CPP-GMR spin valve we consider is schematically shown. 
    The symbols are defined in the text.
  }
  \vspace{-2ex}
  \label{fig:fig1}
\end{figure}


However, recently, Chen \etal \cite{chen06} reported that 
the critical current of STT-driven magnetization dynamics of a CPP-GMR
spin valve remains finite even in the zero-thickness limit of the free layer. 
What are missing from the above consideration based on Slonczewski's
theory are the effects of the finite penetration depth of the transverse spin
current, $\lambda_{\rm t}$,  \cite{zhang02,zhang04,taniguchi08} and spin pumping
\cite{taniguchi08,mizukami02a,mizukami02b,tserkovnyak02,tserkovnyak03}.
The penetration depth of the transverse spin current is the
characteristic length of the
ferromagnetic metal over which the transfer of the spin angular
 momentum from conducting electrons to the magnetization is achieved.  
 If the free layer is thinner than $\lambda_{\rm t}$, the conducting
 electrons cannot
 transfer their angular momentum to the magnetization to exert STT.
Spin pumping is the phenomenon 
by which the spin current is pumped
out from the free layer into the other layers. 
The magnetic (Gilbert)
damping of the free layer is enhanced by spin pumping.
Therefore, we need to analyze the experimental results by taking into
account both the finite penetration depth of the transverse spin current and
spin pumping to understand the mechanism that determines the critical
current of STT-driven magnetization dynamics in magnetic multilayers.


In this paper, we study the critical current of STT-driven
magnetization dynamics 
by taking into account both the finite penetration depth of the
transverse spin current and spin pumping.
In order to analyze the experiments by Chen \etal \cite{chen06}, 
we extend the spin pumping theory with the finite penetration depth
\cite{taniguchi08} to include the electric current.
We show that the critical current remains finite even 
in the zero-thickness limit of the free layer, 
which agrees quantitatively well with the results of Ref. \cite{chen06}, 
and that the remaining value is determined mainly by spin pumping. 
We find that we can control the remaining value of the critical current 
by varying the spin diffusion length of the nonmagnetic electrode
adjacent to the free layer.  
The longer the spin diffusion length of
the nonmagnetic electrode, 
the smaller the remaining value of the critical current.


The system we consider is shown in Fig. \ref{fig:fig1}.  
Two ferromagnetic layers (F${}_{1}$ and F${}_{2}$) are sandwiched 
by the nonmagnetic layers N${}_{i}$ $(i \! = \! 1 \! - \! 7)$.  
The F${}_{1}$ and F${}_{2}$ layers correspond to the free and fixed layer, respectively.  
$\mathbf{m}_{k}$ $(k \! = \! 1,2)$ is the unit vector
pointing the direction of the magnetization of the F${}_{k}$ layer.
$d_{k}$ and $L_{i}$ are the thicknesses of the F${}_{k}$ and N${}_{i}$ layers, respectively. 
The electric current $I$ flows from 
the N${}_{7}$ layer to the N${}_{1}$ layer.


In order to analyze the STT-driven magnetization dynamics in the 
multilayer system shown in Fig. \ref{fig:fig1}, 
we extend the spin-pumping theory with the finite penetration depth
\cite{taniguchi08} to include the electric current. 
The electric current and pumped spin current at 
the F${}_{k}$/N${}_{i}$ interface (into N${}_{i}$) are 
derived by the circuit theory \cite{brataas01},
and expressed in terms of the charge accumulation $\mu_{{\rm N}_{i},{\rm F}_{k}}$  and
the spin accumulation $\bm{\mu}_{{\rm N}_{i},{\rm F}_{k}}$ as 
\cite{tserkovnyak02,brataas01}
\begin{align}
  &
I^{{\rm F}_{k}/{\rm N}_{i}}
   \! = \! 
  \frac{eg}{2h}
  \left\{
    2(\mu_{{\rm F}_{k}}-\mu_{{\rm N}_{i}})
     \! + \! 
    p\mathbf{m}_{k} \! \cdot \! (\bm{\mu}_{{\rm F}_{k}}-\bm{\mu}_{{\rm N}_{i}})
  \right\},
  \label{eq:current}
  \\
  &
  \mathbf{I}_{s}^{\rm pump}
  =
  \frac{\hbar}{4\pi}
  \left(
    g_{\rm r}^{\uparrow\downarrow}
    \mathbf{m}_{1} \! \times \! \frac{\D\mathbf{m}_{1}}{\D t}
     + 
    g_{\rm i}^{\uparrow\downarrow}
    \frac{\D\mathbf{m}_{1}}{\D t}
  \right), 
  \label{eq:pump_current}
\end{align}
where 
$h \! = \! 2\pi\hbar$ is the Planck constant, 
$g \! = \! g^{\uparrow\uparrow} \! + \! g^{\downarrow\downarrow}$ is 
the sum of the spin-up and spin-down conductances, 
$p \! = \! (g^{\uparrow\uparrow} \! - \! g^{\downarrow\downarrow})/(g^{\uparrow\uparrow} \! + \! g^{\downarrow\downarrow})$ 
is the spin polarization of the conductances 
and $g_{\rm r(i)}$ is the real (imaginary) part of the mixing conductance. 
The spin current at each F${}_{k}$/N${}_{i}$ and N${}_{i}$/N${}_{j}$ interface (into N${}_{i}$) are 
given by \cite{taniguchi08,brataas01}
\begin{align}
&
\begin{aligned}
  \!\!\!\mathbf{I}_{s}^{{\rm F}_{k}/{\rm N}_{i}}
   = 
  &\frac{1}{4\pi}
   \! 
  \left[
    g\left\{ \! 
      p(\mu_{{\rm F}_{k}} \! - \! \mu_{{\rm N}_{i}})
       \! + \! 
      \frac{1}{2}\mathbf{m}_{k} \! \cdot \! (\bm{\mu}_{{\rm F}_{k}} \! - \! \bm{\mu}_{{\rm N}_{i}})
     \! \right\} \! 
    \mathbf{m}_{k}
  \right.
\\
  &- \! 
  g_{\rm r}^{\uparrow\downarrow}
  \mathbf{m}_{k} \! \times \! (\bm{\mu}_{{\rm N}_{i}} \! \times \! \mathbf{m}_{k})
   \! - \! 
  g_{\rm i}^{\uparrow\downarrow}
  \bm{\mu}_{{\rm N}_{i}} \! \times \! \mathbf{m}_{k}
\\
  &+ \!  \left.
   t_{\rm r}^{\uparrow\downarrow}
   \mathbf{m}_{k} \! \times \! (\bm{\mu}_{{\rm F}_{k}} \! \times \! \mathbf{m}_{k})
    \! + \! 
   t_{\rm i}^{\uparrow\downarrow}
   \bm{\mu}_{{\rm F}_{k}} \! \times \! \mathbf{m}_{k}
  \right],
  \label{eq:spin_current_NF}
\end{aligned}
\\
&
  \mathbf{I}_{s}^{{\rm N}_{i}/{\rm N}_{j}}
  =
  -\frac{g_{{\rm N}_{i}/{\rm N}_{j}}}{4\pi}
  (\bm{\mu}_{{\rm N}_{i}} \! - \! \bm{\mu}_{{\rm N}_{j}}),
  \label{eq:spin_current_NN}
\end{align}
where 
$t_{\rm r(i)}^{\uparrow\downarrow}$ is the real (imaginary) part of 
the transmission mixing conductance at the F${}_{k}$/N${}_{i}$ interface 
and $g_{{\rm N}_{i}/{\rm N}_{j}}$ is 
the conductance of the one spin channel at the interface. 
The spin current of Eq. (\ref{eq:spin_current_NF}) is obtained 
from the circuit theory of Brataas \etal \cite{brataas01} 
eliminating the assumption that the non-equilibrium distribution function 
of the electrons in a ferromagnetic layer is aligned to 
the direction of the magnetization in spin space. 
It should be noted that 
the transmission mixing conductance in Eq. (\ref{eq:spin_current_NF}) is 
different from that defined by Zwierzycki \etal \cite{zwierzycki05}.
Zwierzycki \etal calculated 
a transmission mixing conductance defined through a N/F/N junction 
defined as $t^{\uparrow\downarrow}=t^{\uparrow}t^{\downarrow*}$ 
where $t^{\sigma}=t_{{\rm F}\to{\rm N}}^{\sigma}\E^{\I k_{\perp}^{\sigma}d}t_{{\rm N}\to{\rm F}}^{\sigma}$ 
and $t_{{\rm F(N)}\to{\rm N(F)}}^{\sigma}$ is the tranmission coefficient 
for electrons from F (N) to N (F), 
and showed that $t^{\uparrow\downarrow}$ depends on the thickness of 
the ferromagnetic layer $d$ due to the phase factor $\E^{\I k_{\perp}^{\sigma}d}$ \cite{zwierzycki05}. 
On the other hand, 
the transmission mixing conductance in Eq. (\ref{eq:spin_current_NF}) is defined by 
$t_{\rm r(i)}^{\uparrow\downarrow}=\Re(\Im)[t_{{\rm F}\to{\rm N}}^{\uparrow}t_{{\rm F}\to{\rm N}}^{\downarrow *}]$, 
and independent of the thickness of the ferromagnetic layer. 
Although the original formulation of the circuit theory assumed 
the spatially uniform charge and spin accumulation \cite{brataas01}, 
it has been shown that the circuit theory is applicable 
to the diffusive system \cite{tserkovnyak03,bauer03}. 
It should be noted that there is a controversial issue 
regarding the transverse spin accumulation 
in the ferromagnetic layer, 
$\bm{\mu}_{\rm F}^{\rm T} \! = \! \mathbf{m} \! \times \!
(\bm{\mu}_{\rm F} \! \times \! \mathbf{m})$ 
\cite{zhang02,zhang04,tserkovnyak02,tserkovnyak03,brataas01,slonczewski06,urazhdin05,guo05}.


The spin accumulation in the nonmagnetic layer, 
$\bm{\mu}_{\rm N}$ obeys the diffusion equation \cite{valet93}, 
and is expressed as a linear combination of 
$\exp(\pm x/\lambda_{\rm sd(N)})$, 
where $\lambda_{\rm sd(N)}$ is the spin diffusion length of the nonmagnetic layer. 
The spin current in the nonmagnetic layer is given by 
\begin{equation}
  \mathbf{I}_{s}^{\rm N}
  =
  -\frac{\partial}{\partial x}
  \frac{\hbar S\sigma_{\rm N}}{2e^{2}}
  \bm{\mu}_{\rm N},
  \label{eq:spin_current_N}
\end{equation}
where $\sigma_{\rm N}$ is the conductivity of the nonmagnetic layer. 


The longitudinal spin accumulation in the ferromagnetic layer, 
$\bm{\mu}_{\rm F}^{\rm L} \! = \! (\mathbf{m} \! \cdot \!
\bm{\mu}_{\rm F})\mathbf{m}$, 
also satisfies the diffusion equation, 
and is expressed as a linear combination of $\exp(\pm x/\lambda_{\rm sd(F_{L})})$, 
where $\lambda_{\rm sd(F_{L})}$ is the longitudinal spin diffusion length 
of the ferromagnetic layer \cite{valet93}. 
The longitudinal spin current in the ferromagnetic layer is 
\begin{equation}
  (\mathbf{m}\cdot\mathbf{I}_{s}^{\rm F})\mathbf{m}
  =
  -\frac{\partial}{\partial x}
  \frac{\hbar S}{2e^{2}}
  (\sigma_{\rm F}^{\uparrow}\mu_{\rm F}^{\uparrow}-\sigma_{\rm F}^{\downarrow}\mu_{\rm F}^{\downarrow})
  \mathbf{m},
  \label{eq:spin_current_FL}
\end{equation}
where $\mu_{\rm F}^{\uparrow(\downarrow)} \!\! = \!\! \int_{\varepsilon_{\rm F}}\!\D\varepsilon f^{\uparrow(\downarrow)}$ 
and $\sigma_{\rm F}^{\uparrow(\downarrow)}$ are 
the electro-chemical potential and the conductivity for the spin-up (spin-down) electrons, respectively. 
The spin polarization of the conductivity is defined as
$\beta \! = \! (\sigma_{\rm F}^{\uparrow}-\sigma_{\rm F}^{\downarrow})/(\sigma_{\rm F}^{\uparrow}+\sigma_{\rm F}^{\downarrow})$.


The transverse spin accumulation in the ferromagnetic layer 
obeys \cite{zhang02}
\begin{equation}
  \frac{\partial^{2}}{\partial x^{2}}
  \bm{\mu}_{\rm F}^{\rm T}
  =
  \frac{1}{\lambda_{J}^{2}}
  \bm{\mu}_{\rm F}^{\rm T} \! \times \! \mathbf{m}
  +
  \frac{1}{\lambda_{\rm sd(F_{T})}^{2}}
  \bm{\mu}_{\rm F}^{\rm T},
  \label{eq:diff_eq_FT}
\end{equation}
where $\lambda_{J} \! = \! \sqrt{(D_{\rm F}^{\uparrow}+D_{\rm F}^{\downarrow})\hbar/(2J)}$ 
and $\lambda_{\rm sd(F_{T})}$ is the transverse spin diffusion length. 
$J$ is the strength of the exchange field \cite{zhang04}
and $D_{\rm F}^{\uparrow(\downarrow)}$ is 
the diffusion constant of spin-up (spin-down) electrons. 
The spin polarization of the diffusion constant is defined as
$\beta' \! = \! (D_{\rm F}^{\uparrow} \! - \! D_{\rm F}^{\downarrow})/(D_{\rm F}^{\uparrow} \! + \! D_{\rm F}^{\downarrow})$. 
The transverse spin accumulation is expressed as
a linear combination of $\exp(\pm x/l_{+})$ and $\exp(\pm x/l_{-})$, 
where $1/l_{\pm} \! = \! \sqrt{(1/\lambda_{\rm sd(F_{T})}^{2}) \! \mp \! (\I/\lambda_{J}^{2})}$. 
The penetration depth of the transverse spin current 
$\lambda_{\rm t}$ is defined as
$1/\lambda_{\rm t} \! = \! \Re[1/l_{+}]$ \cite{taniguchi08}.
The transverse spin current in the ferromagnetic layer is expressed as
\begin{equation}
  \mathbf{m} \! \times \! (\mathbf{I}_{s}^{\rm F} \! \times \! \mathbf{m})
  =
  -\frac{\partial}{\partial x}
  \frac{\hbar S\sigma_{\rm F}^{\uparrow\downarrow}}{2e^{2}}
  \bm{\mu}_{\rm F}^{\rm T},
  \label{eq:spin_current_FT}
\end{equation}
where 
$\sigma_{\rm F}^{\uparrow\downarrow} \! 
  = \! [\sigma_{\rm F}^{\uparrow}/(1 \! + \! \beta') \! + \! \sigma_{\rm F}^{\downarrow}/(1 \! - \! \beta')]/2$ 
\cite{zhang02,taniguchi08}.


The total spin currents 
across the N${}_{3}$/F${}_{1}$ and F${}_{1}$/N${}_{4}$ interfaces, i.e., 
$\mathbf{I}_{s}^{(1)} \! = \! \mathbf{I}_{s}^{\rm pump} \! + \! \mathbf{I}_{s}^{\rm F_{1}/N_{3}}$ and 
$\mathbf{I}_{s}^{(2)} \! = \! \mathbf{I}_{s}^{\rm pump} \! + \!
\mathbf{I}_{s}^{\rm F_{1}/N_{4}}$, exert the torque 
$\bm{\tau} \!=\! 
  \mathbf{m}_{1} \! \times \![(\mathbf{I}_{s}^{(1)}\!+\!\mathbf{I}_{s}^{(2)})\!\times\!\mathbf{m}_{1}]$ 
on the magnetization $\mathbf{m}_{1}$. 
In order to obtain the spin current $\mathbf{I}_{s}^{(1,2)}$,  we
solve the diffusion equations of spin accumulations in each layer. 
The boundary conditions  are as follows.
We assume that the thickness of the N${}_{1}$ and N${}_{7}$ layer, $L_{1}$ and $L_{7}$, 
are sufficiently thick enough compared to their spin diffusion length, 
and that the spin current is zero at the outer boundary of the N${}_{1}$ and N${}_{7}$ layer. 
We also assume that the spin current is continuous at all F${}_{k}$/N${}_{i}$ and
N${}_{i}$/N${}_{j}$ interfaces 
and that the electric current is constant 
through the entire structure. 
The spin current $\mathbf{I}_{s}^{(1,2)}$ is obtained as a function of 
the electric current $I$ and the pumped spin current
$\mathbf{I}_{s}^{\rm pump}$.


The torque $\bm{\tau}$ modifies 
the Landau-Lifshitz-Gilbert (LLG) equation 
of the magnetization $\mathbf{m}_{1}$. 
The LLG equation conserves the magnitude of the magnetization, 
and thus the vectors 
$\dot{\mathbf{m}}_{1}$ and $\mathbf{m}_{1}\!\!\times\!\dot{\mathbf{m}}_{1}$ are 
perpendicular to the magnetization $\mathbf{m}_{1}$. 
Since the torque $\bm{\tau}$ is perpendicular to $\mathbf{m}_{1}$ 
the torque can be decomposed into the directions of
$\dot{\mathbf{m}}_{1}$ and $\mathbf{m}_{1}\!\!\times\!\dot{\mathbf{m}}_{1}$. 
The LLG equation of $\mathbf{m}_{1}$ is expressed as
\cite{tserkovnyak02,taniguchi08,taniguchi07}
\begin{equation}
\begin{split}
  \frac{\D\mathbf{m}_{1}}{\D t}
   \! &= \! 
  -\gamma\mathbf{m}_{1} \! \times \! \mathbf{B}_{\rm eff}
  +
  \frac{\gamma}{MSd_{1}}\bm{\tau}
  +
  \alpha_{0}\mathbf{m}_{1} \! \times \! \frac{\D\mathbf{m}_{1}}{\D t}
\\
  &=
  -\gamma_{\rm eff}\mathbf{m}_{1} \! \times \! \mathbf{B}_{\rm eff}
  +
  \frac{\gamma_{\rm eff}}{\gamma}
  (\alpha_{0}+\alpha')
  \mathbf{m}_{1} \! \times \! \frac{\D\mathbf{m}_{1}}{\D t},
  \label{eq:LLG}
\end{split}
\end{equation}
where $\mathbf{B}_{\rm eff}$ is the effective magnetic field. 
$\alpha'=\alpha_{c}+\alpha_{\rm pump}$ represents the enhancement of 
the Gilbert damping constant. 
The enhancement $\alpha_{c}$ is proportional to the electric current $I$ 
and independent of the pumped spin current $\mathbf{I}_{s}^{\rm pump}$. 
The enhancement $\alpha_{\rm pump}$ represents the contribution from
the pumped spin current and is independent of the electric current. 
It should be noted that 
the enhancement $\alpha_{\rm pump}$ differs from 
the result of the conventional spin-pumping theory \cite{tserkovnyak03} 
because $\alpha_{\rm pump}$ is a function of $\lambda_{\rm t}$.
The enhancement of 
the gyromagnetic ratio $\gamma_{\rm eff}/\gamma$ 
is also a function of the electric current and 
the pumped spin current.

Let us move to the analysis of experimental results of 
Ref. \cite{chen06}.  In general, 
the dynamics of the magnetization $\mathbf{m}_{1}$ 
determined by Eq. (\ref{eq:LLG}) is very complicated; 
thus, we cannot obtain the analytical expression of the critical current of 
STT-driven magnetization dynamics
of the magnetization $\mathbf{m}_{1}$. 
However, in the experiment of Ref. \cite{chen06}, 
the system, and therefore the dynamics of $\mathbf{m}_{1}$, 
have axial symmetry along the direction normal to the film plane 
because the high magnetic field (about 7 T) is applied along this direction. 
Then we assume that 
the magnetization of the F${}_{1}$ layer $\mathbf{m}_{1}$ precesses around 
the magnetization of the F${}_{2}$ layer $\mathbf{m}_{2}$ 
with the relative angle $\theta$ and the angular frequency $\omega$. 
The critical current of STT-driven magnetization dynamics is defined by 
the current that satisfies the condition, 
$\alpha_{0}+\alpha_{c}+\alpha_{\rm pump}=0$. 
The critical current $I_{\rm c}$ is expressed as
\begin{equation}
  I_{\rm c}
  =
  \frac{2eMSd_{1}}{\hbar\gamma \tilde{\eta}}
  (\alpha_{0}+\alpha_{\rm pump})\omega,
  \label{eq:critical_current2}
\end{equation}
where $\tilde{\eta}$ is the effective transverse spin polarization coefficient 
that is determined by the diffusion equations of 
the spin accumulations,
and thus the coefficient 
$\tilde{\eta}$ is the function of $d_{1}/\lambda_{\rm sd(F_{L})}$ and
$d_{1}/l_{\pm}$.

The parameters we used are as follows. 
The system consists of nine layers shown in Fig. \ref{fig:fig1}, 
where F${}_{1}$ and F${}_{2}$ are Co, 
N${}_{1}$, N${}_{3}$, N${}_{4}$, N${}_{5}$ and N${}_{7}$ are Cu, 
and N${}_{2}$ and N${}_{6}$ are Pt. 
The thicknesses of the N${}_{3}$, N${}_{4}$ and N${}_{5}$ layers are 10 nm, 
the thicknesses of the N${}_{2}$ and N${}_{6}$ layers are 3 nm 
and the thickness of the F${}_{2}$ layer is 12 nm \cite{chen06}. 
The thickness of the N${}_{1}$ and N${}_{7}$ layers are taken to be 10 $\mu$m, 
which is sufficiently longer than the spin diffusion length. 
The resistivity $(2\sigma_{\rm N})^{-1}$ of Cu and Pt are 
14 and 42 $\Omega$nm, respectively \cite{bass07}. 
The spin diffusion length $\lambda_{\rm N}$ of Cu and Pt are 
1000 and 14 nm, respectively \cite{bass07}. 
The conductance at the Cu/Pt interface 
$g_{\rm Cu/Pt}/S$ is 35 nm${}^{-2}$ \cite{bass07}. 
The magnetization $M$, 
the intrinsic Gilbert damping constant $\alpha_{0}$ and
the gyromagnetic ratio $\gamma$ of Co are 
0.14 T, 0.008 and $1.89 \! \times \! 10^{11}$ Hz/T, 
respectively \cite{chen06,beaujour06}. 
For simplicity, 
we assume that $p$=$\beta$=$\beta^{'}$=0.46 for Co \cite{bass07}. 
The resistivity $(\sigma_{\rm F}^{\uparrow} \! + \! \sigma_{\rm F}^{\downarrow})^{-1}$ 
and the longitudinal spin diffusion length $\lambda_{\rm sd(F_{L})}$ 
of the Co are 60 $\Omega$nm and 40 nm, respectively \cite{bass07}. 
The transverse spin diffusion length is 
$\lambda_{\rm sd(F_{T})} \! = \! \lambda_{\rm sd(F_{L})}/\sqrt{1 \! - \! \beta^{2}}$ \cite{zhang02}. 
$\lambda_{J}$ is taken to be 3.0 nm \cite{zhang04}, 
i.e., $\lambda_{\rm t} \! = \! 4.2$ nm. 
The conductances at the Co/Cu interface, 
$g/S$, $g_{\rm r}^{\uparrow\downarrow}/S$ and $g_{\rm i}^{\uparrow\downarrow}/S$, are 
50, 27 and 0.4 nm${}^{-2}$, respectively \cite{tserkovnyak02,tserkovnyak03,brataas01,zwierzycki05,xia02}. 
The angular frequency is 
$\omega \! = \! \gamma(B_{\rm appl} \! + \! 4\pi M)$ 
where the strength of the applied magnetic field 
$B_{\rm appl}$ is 7 T \cite{chen06}. 
The relative angle of the two magnetizations $\theta$ 
is assumed to be $0.99\pi$ \cite{chen06}. 
Although there are many material parameters in our calculation 
these values except $t_{\rm r,i}^{\uparrow\downarrow}$ are determined by 
the experiments and \textit{ab initio} calculations. 
The value of $t_{\rm r,i}^{\uparrow\downarrow}/S$ is determied by fitting, 
and taken to be 6.0 nm${}^{-2}$. 
According to Ref. \cite{chen06}, 
the experimental values are the low temperature values. 


\begin{figure}
  \centerline{\includegraphics[width=0.8\columnwidth]{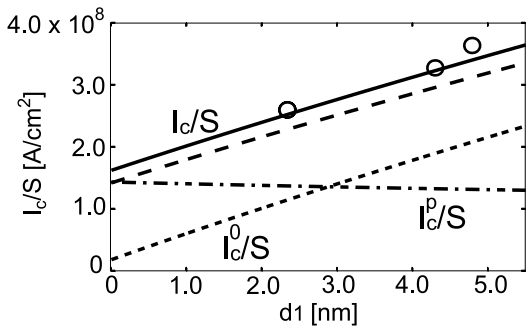}}
  \caption{
  The critical current density vs 
  the thickness of the free (F${}_{1}$) layer. 
  The circles are the experimental result of Chen \etal 
  for $70 \! \times \! 140$ nm${}^{2}$ junctions \cite{chen06}.
  The solid line 
  corresponds to $I_{\rm c}/S$ (see Eq. (\ref{eq:critical_current2}).) 
  The dotted line 
  and dashed-dotted line 
  correspond to 
  $I_{\rm c}^{0}/S$ and $I_{\rm c}^{\rm p}/S$, respectively. 
  The dashed line corresponds to $I_{\rm c}/S$ 
  in the limit of $\lambda_{\rm t}\to 0$. 
  }
  \vspace{-3ex}
  \label{fig:fig2}
\end{figure}


The obtained critical current density is plotted by a solid line
against the thickness of the free layer $d_{1}$ in
Fig. \ref{fig:fig2}.  
The experimental results of Ref. \cite{chen06} are shown by open circles.  
One can see that our results 
agree well with the experimental results. 
The critical current density decreases 
as the thickness of the free layer decreases, and
remains finite even in the zero-thickness limit of the free layer. 
In order to see the main mechanism that determines the remaining
value of the critical current density, we decompose $I_{\rm c}$ of 
Eq. (\ref{eq:critical_current2}) into two parts as 
$I_{\rm c}\!=\!I_{\rm c}^{0}\!+\!I_{\rm c}^{\rm p}$,
where $I_{\rm c}^{0}$ is the component proportional to $\alpha_{0}$ 
and $I_{\rm c}^{\rm p}$ is the component proportional to $\alpha_{\rm pump}$. 
In Fig. \ref{fig:fig2}, the components $I_{\rm c}^{0}/S$ and $I_{\rm c}^{\rm p}/S$ 
are plotted by dotted and dot-dashed lines, respectively.
As shown in Fig. \ref{fig:fig2} the remaining value of the critical
current in the limit of $d_{1}\to 0$ is determined mainly by the spin pumping. 
Although $I_{\rm c}^{0}/S$ is also finite in the limit of $d_{1}\to 0$ 
because of the finite penetration depth of the transverse spin current 
$\lambda_{\rm t}$ in the F${}_{1}$ layer, 
the remaining value is small compared to $I_{\rm c}^{\rm p}/S$. 
The dashed line in Fig. \ref{fig:fig2} is 
the calculated critical current $I_{\rm c}/S$ in the limit of $\lambda_{\rm t}\to 0$.
According to Fig. \ref{fig:fig2} 
we conclude that the effect of the finite penetration depth $\lambda_{\rm t}$ is 
less important to describe the results of Ref. \cite{chen06}.


The reason why both $I_{\rm c}^{0}$ and $I_{\rm c}^{\rm P}$ remain finite 
in the limit of $d_{1}\to 0$ is understood as follows. 
Slonczewski assumed that the transverse spin current 
injected into the free layer is absorbed at the interface, 
and thus, STT is independent of the thickness of the free layer. 
The critical current is determined by the competition 
between STT and the magnetic (Gilbert) damping 
of the free layer. 
The spin relaxation due to the Gilbert damping is 
proportional to the thickness of the free layer $d_{1}$, 
and thus, the critical current given by Eq. (\ref{eq:critical_current1}) 
is proportional to $d_{1}$ and vanishes in the limit of $d_{1}\to 0$. 
If the penetration depth of the transverse spin current $\lambda_{\rm t}$ is finite, 
the transverse spin current is not fully absorbed in the free layer 
in the case of $d_{1}\ll \lambda_{\rm t}$. 
Then the strength of STT is decreased compared to 
the prediction of Slonczewski \cite{slonczewski96}, 
and thus, the critical current is increased. 
Spin pumping enhances the Gilbert damping, 
and the spin relaxation due to spin pumping is 
independent of the thickness of the free layer. 
Thus, $I_{\rm c}^{\rm p}$ remains finite 
in the limit of $d_{1}\to 0$.


The dependences of the remaining value, 
about $1.6\times 10^{8}$ A/cm${}^{2}$, 
on the parameters given above are as follows. 
If the resistivity and the longitudinal spin diffusion length 
of Co are taken to be 210 $\Omega$nm and 38 nm, respectively, 
which are the room temperature values \cite{bass07}, 
the remaining value is estimated to be $1.5\times 10^{8}$ A/cm${}^{2}$. 
The reduction the longitudinal spin diffusion length decreases
the penetration depth $\lambda_{\rm t}$, 
and thus, the remaining value is reduced. 
The values of conductances, $g$ and $g_{\rm r,i}^{\uparrow\downarrow}$, 
include the effect of the Sharvin conductace \cite{zwierzycki05}. 
If $g/S$, $g_{\rm r}^{\uparrow\downarrow}/S$ and $g_{\rm i}^{\uparrow\downarrow}/S$
are taken to be 19.3, 14.6 and -1.1 nm${}^{-2}$, respectively, 
which are the bare values estimated by $\textit{ab initio}$ calculation \cite{zwierzycki05},
the remaining value is estimated to be $1.0\times 10^{8}$ A/cm${}^{2}$. 
The reduction of the mixing conductance decreases the effect 
of spin pumping \cite{tserkovnyak02}, 
and thus, the remaining value is reduced. 
If the transmission mixing conductance $t_{\rm r,i}^{\uparrow\downarrow}/S$ is 
taken to be 3.0 (12.0) nm${}^{-2}$, 
which is half (twice) compared to the value used in Fig. \ref{fig:fig2}, 
the remaining value is estimated to be $1.5\ (1.8)\times 10^{8}$ A/cm${}^{2}$. 
The reduction (enhancement) of the transmission mixing conductance 
decreases (increases) the effect of the transverse spin accumulation, 
or equivallently the penetration depth,
on the spin current given by Eq. (\ref{eq:spin_current_NF}), 
and thus, the remaining value is reduced (enhanced). 
We conclude that although there are many material parameters in our calculation 
the parameter dependence of the remaining value is small, 
and our calculation gives the correct order of the critical current.

The above results imply that we can increase or decrease the
critical current by controlling the spin pumping.
Spin pumping is the phenomenon by which the precessing magnetization of the
free layer pumps spin current into the other layers. 
The other layers act as an additional spin sink and 
the magnetic damping of the free layer is enhanced by spin pumping.  
The ability of the spin sink is determined by the spin diffusion length
since the spin diffusion length is inversely proportional to the square root of
the spin scattering rate.  
Materials with short (long) spin
diffusion length act as a good (bad) spin sink.
One may expect that if the nonmagnetic layer adjacent to the free
layer is made of material with a long spin diffusion length, 
the Gilbert damping constant and therefore the critical current is
suppressed.
In the limit of infinite spin diffusion length, $\lambda_{\rm
  N}\to\infty$, there is no spin flip scattering in the nonmagnetic
layer and the spin pumping into the nonmagnetic layer is forbidden.
On the other hand, if the nonmagnetic layer adjacent to the free
layer is made of a material with a short spin diffusion length, 
the Gilbert damping constant and the critical current is enhanced.
In the limit of $\lambda_{\rm N} \to 0$,
the pumped spin current is absorbed at the interface and enhancement
of the critical current due to spin current is maximized. 

\begin{figure}
  \centerline{\includegraphics[width=0.8\columnwidth]{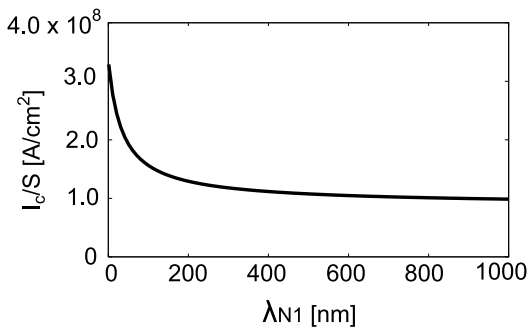}}
  \caption{
  The dependence of the critical current 
  in the zero-thickness limit of the free layer 
  in the N${}_{1}$/F${}_{1}$/N${}_{4}$/F${}_{2}$/N${}_{7}$ 
  five-layer system 
  on the spin diffusion length of the N${}_{1}$ layer, $\lambda_{\rm N_{1}}$. 
  }
  \vspace{-3ex}
  \label{fig:fig3}
\end{figure}

In order to verify the above statement, we analyzed the critical
current of the five-layer system, 
N${}_{1}$/F${}_{1}$/N${}_{4}$/F${}_{2}$/N${}_{7}$ (see Fig. \ref{fig:fig1}), 
where all parameters except the spin diffusion length of the N${}_{1}$
layer, $\lambda_{\rm N_{1}}$, are the same as those used in the analysis
of Chen's experiment.
In Fig. \ref{fig:fig3}, we plot the critical current in the
zero-thickness limit of the free layer as a function of the spin
diffusion length of the N${}_{1}$ layer $\lambda_{{\rm N}_{1}}$. 
One can see that the critical current is a decreasing function of
$\lambda_{{\rm N}_{1}}$. 
The critical current remains finite in the limit of
$\lambda_{\rm N}\to\infty$ 
because of the spin pumping into the N${}_{4}$ layer and 
the finite penetration depth of the transverse spin current. 
The result shown in Fig. \ref{fig:fig3} shows that
we can control the critical current by varying the spin diffusion length of
the nonmagnetic electrode adjacent to the free layer.

We cannot apply the present formula directly to a magnetic tunnel
junction (MTJ) because spin accumulation is not well-defined in an
insulator (I).  Although the spin pumping across the insulating barrier
is beyond the scope of this paper, the spin pumping into the metallic
electrode should give the finite remaining value of the critical
current. 
Recently spin pumping in a magnetic tunnel junction (MTJ) 
was studied by Moriyama \etal \cite{moriyama08}. 
They studied a ferromagnetic resonance
in Al/AlO/Ni${}_{80}$Fe${}_{20}$/Cu MTJ, 
and found the generation of the voltage on the order of few $\mu$V, 
which is qualitatively explained by 
the theory of spin pumping in a metallic structure \cite{wang06}, 
but requires unreasonably large value of mixing conductance. 
The results of Moriyama \etal suggest that a new non-equilibtirum
phenomena exists in MTJ, e.g. charge pumping or the development of the
longitudinal spin accumulation in a ferromagnetic layer
\cite{tserkovnyak08}.

In conclusion, we studied the critical current of 
spin transfer torque-driven magnetization dynamics 
by taking into account both the finite penetration depth 
of the transverse spin current in the ferromagnetic
layer and spin pumping.
We extend the spin-pumping theory with the finite penetration depth
to include the electric current and successfully reproduced the 
experimental results of Ref. \cite{chen06}. 
We showed that the critical current remains finite
in the zero-thickness limit of the free layer and the remaining value is
determined mainly by spin pumping.
We also showed we can control the remaining value of the critical current 
by varying the spin diffusion length of the nonmagnetic electrode
adjacent to the free layer.


The authors would like to thank W. Chen, A. D. Kent and their co-workers 
for providing their experimental results.
This work was supported by JSPS and NEDO.



\begin{thebibliography}{38}
\expandafter\ifx\csname natexlab\endcsname\relax\def\natexlab#1{#1}\fi
\expandafter\ifx\csname bibnamefont\endcsname\relax
  \def\bibnamefont#1{#1}\fi
\expandafter\ifx\csname bibfnamefont\endcsname\relax
  \def\bibfnamefont#1{#1}\fi
\expandafter\ifx\csname citenamefont\endcsname\relax
  \def\citenamefont#1{#1}\fi
\expandafter\ifx\csname url\endcsname\relax
  \def\url#1{\texttt{#1}}\fi
\expandafter\ifx\csname urlprefix\endcsname\relax\def\urlprefix{URL }\fi
\providecommand{\bibinfo}[2]{#2}
\providecommand{\eprint}[2][]{\url{#2}}

\bibitem[{\citenamefont{Slonczewski}(1996)}]{slonczewski96}
\bibinfo{author}{\bibfnamefont{J.~C.} \bibnamefont{Slonczewski}},
  \bibinfo{journal}{J. Magn. Magn. Mater.} \textbf{\bibinfo{volume}{159}},
  \bibinfo{pages}{L1} (\bibinfo{year}{1996}).

\bibitem[{\citenamefont{Berger}(1996)}]{berger96}
\bibinfo{author}{\bibfnamefont{L.}~\bibnamefont{Berger}},
  \bibinfo{journal}{Phys. Rev. B} \textbf{\bibinfo{volume}{54}},
  \bibinfo{pages}{9353} (\bibinfo{year}{1996}).

\bibitem[{\citenamefont{Katine et~al.}(2000)\citenamefont{Katine, Albert,
  Buhrman, Myers, and Ralph}}]{katine00}
\bibinfo{author}{\bibfnamefont{J.~A.} \bibnamefont{Katine}},
  \bibinfo{author}{\bibfnamefont{F.~J.} \bibnamefont{Albert}},
  \bibinfo{author}{\bibfnamefont{R.~A.} \bibnamefont{Buhrman}},
  \bibinfo{author}{\bibfnamefont{E.~B.} \bibnamefont{Myers}}, \bibnamefont{and}
  \bibinfo{author}{\bibfnamefont{D.~C.} \bibnamefont{Ralph}},
  \bibinfo{journal}{Phys. Rev. Lett.} \textbf{\bibinfo{volume}{84}},
  \bibinfo{pages}{3149} (\bibinfo{year}{2000}).

\bibitem[{\citenamefont{Carva et~al.}(2006)\citenamefont{Carva, Turek,
  Kudrnovsk\'y, and Bengone}}]{carva06}
\bibinfo{author}{\bibfnamefont{K.}~\bibnamefont{Carva}},
  \bibinfo{author}{\bibfnamefont{I.}~\bibnamefont{Turek}},
  \bibinfo{author}{\bibfnamefont{J.}~\bibnamefont{Kudrnovsk\'y}},
  \bibnamefont{and} \bibinfo{author}{\bibfnamefont{O.}~\bibnamefont{Bengone}},
  \bibinfo{journal}{Phys. Rev. B} \textbf{\bibinfo{volume}{73}},
  \bibinfo{pages}{144421} (\bibinfo{year}{2006}).

\bibitem[{\citenamefont{Haney et~al.}(2007)\citenamefont{Haney, Waldron, Duine,
  N\'unez, Guo, and MacDonald}}]{haney07}
\bibinfo{author}{\bibfnamefont{P.~M.} \bibnamefont{Haney}},
  \bibinfo{author}{\bibfnamefont{D.}~\bibnamefont{Waldron}},
  \bibinfo{author}{\bibfnamefont{R.~A.} \bibnamefont{Duine}},
  \bibinfo{author}{\bibfnamefont{A.~S.} \bibnamefont{N\'unez}},
  \bibinfo{author}{\bibfnamefont{H.}~\bibnamefont{Guo}}, \bibnamefont{and}
  \bibinfo{author}{\bibfnamefont{A.~H.} \bibnamefont{MacDonald}},
  \bibinfo{journal}{Phys. Rev. B} \textbf{\bibinfo{volume}{76}},
  \bibinfo{pages}{024404} (\bibinfo{year}{2007}).

\bibitem[{\citenamefont{Heiliger and Stiles}(2008)}]{heiliger08}
\bibinfo{author}{\bibfnamefont{C.}~\bibnamefont{Heiliger}} \bibnamefont{and}
  \bibinfo{author}{\bibfnamefont{M.~D.} \bibnamefont{Stiles}},
  \bibinfo{journal}{Phys. Rev. Lett.} \textbf{\bibinfo{volume}{100}},
  \bibinfo{pages}{186805} (\bibinfo{year}{2008}).

\bibitem[{\citenamefont{Kubota et~al.}(2008)\citenamefont{Kubota, Fukushima,
  Yakushiji, Nagahama, Yuasa, Ando, Maehara, Nagamine, Tsunekawa, Djayaprawira
  et~al.}}]{kubota08}
\bibinfo{author}{\bibfnamefont{H.}~\bibnamefont{Kubota}},
  \bibinfo{author}{\bibfnamefont{A.}~\bibnamefont{Fukushima}},
  \bibinfo{author}{\bibfnamefont{K.}~\bibnamefont{Yakushiji}},
  \bibinfo{author}{\bibfnamefont{T.}~\bibnamefont{Nagahama}},
  \bibinfo{author}{\bibfnamefont{S.}~\bibnamefont{Yuasa}},
  \bibinfo{author}{\bibfnamefont{K.}~\bibnamefont{Ando}},
  \bibinfo{author}{\bibfnamefont{H.}~\bibnamefont{Maehara}},
  \bibinfo{author}{\bibfnamefont{Y.}~\bibnamefont{Nagamine}},
  \bibinfo{author}{\bibfnamefont{K.}~\bibnamefont{Tsunekawa}},
  \bibinfo{author}{\bibfnamefont{D.~D.} \bibnamefont{Djayaprawira}},
  \bibnamefont{et~al.}, \bibinfo{journal}{Nature Phys.}
  \textbf{\bibinfo{volume}{4}}, \bibinfo{pages}{37} (\bibinfo{year}{2008}).

\bibitem[{\citenamefont{Sankey et~al.}(2008)\citenamefont{Sankey, Cui, Sun,
  Slonczewski, Buhrman, and Ralph}}]{sankey08}
\bibinfo{author}{\bibfnamefont{J.~C.} \bibnamefont{Sankey}},
  \bibinfo{author}{\bibfnamefont{Y.-T.} \bibnamefont{Cui}},
  \bibinfo{author}{\bibfnamefont{J.~Z.} \bibnamefont{Sun}},
  \bibinfo{author}{\bibfnamefont{J.~C.} \bibnamefont{Slonczewski}},
  \bibinfo{author}{\bibfnamefont{R.~A.} \bibnamefont{Buhrman}},
  \bibnamefont{and} \bibinfo{author}{\bibfnamefont{D.~C.} \bibnamefont{Ralph}},
  \bibinfo{journal}{Nature Phys.} \textbf{\bibinfo{volume}{4}},
  \bibinfo{pages}{67} (\bibinfo{year}{2008}).

\bibitem[{\citenamefont{Kiselev et~al.}(2003)\citenamefont{Kiselev, Sankey,
  Krivorotov, Emley, Schoelkopf, Buhrman, and Ralph}}]{kiselev03}
\bibinfo{author}{\bibfnamefont{S.~I.} \bibnamefont{Kiselev}},
  \bibinfo{author}{\bibfnamefont{J.~C.} \bibnamefont{Sankey}},
  \bibinfo{author}{\bibfnamefont{I.~N.} \bibnamefont{Krivorotov}},
  \bibinfo{author}{\bibfnamefont{N.~C.} \bibnamefont{Emley}},
  \bibinfo{author}{\bibfnamefont{R.~J.} \bibnamefont{Schoelkopf}},
  \bibinfo{author}{\bibfnamefont{R.~A.} \bibnamefont{Buhrman}},
  \bibnamefont{and} \bibinfo{author}{\bibfnamefont{D.~C.} \bibnamefont{Ralph}},
  \bibinfo{journal}{Nature} \textbf{\bibinfo{volume}{425}},
  \bibinfo{pages}{380} (\bibinfo{year}{2003}).

\bibitem[{\citenamefont{Seki et~al.}(2006)\citenamefont{Seki, Mitani,
  Yakushiji, and Takanashi}}]{seki06}
\bibinfo{author}{\bibfnamefont{T.}~\bibnamefont{Seki}},
  \bibinfo{author}{\bibfnamefont{S.}~\bibnamefont{Mitani}},
  \bibinfo{author}{\bibfnamefont{K.}~\bibnamefont{Yakushiji}},
  \bibnamefont{and}
  \bibinfo{author}{\bibfnamefont{K.}~\bibnamefont{Takanashi}},
  \bibinfo{journal}{Appl. Phys. Lett.} \textbf{\bibinfo{volume}{89}},
  \bibinfo{pages}{172504} (\bibinfo{year}{2006}).

\bibitem[{\citenamefont{Deac et~al.}(2006)\citenamefont{Deac, Lee, Liu, Redon,
  Li, Wang, Nozi\'eres, and Dieny}}]{deac06}
\bibinfo{author}{\bibfnamefont{A.}~\bibnamefont{Deac}},
  \bibinfo{author}{\bibfnamefont{K.~J.} \bibnamefont{Lee}},
  \bibinfo{author}{\bibfnamefont{Y.}~\bibnamefont{Liu}},
  \bibinfo{author}{\bibfnamefont{O.}~\bibnamefont{Redon}},
  \bibinfo{author}{\bibfnamefont{M.}~\bibnamefont{Li}},
  \bibinfo{author}{\bibfnamefont{P.}~\bibnamefont{Wang}},
  \bibinfo{author}{\bibfnamefont{J.~P.} \bibnamefont{Nozi\'eres}},
  \bibnamefont{and} \bibinfo{author}{\bibfnamefont{B.}~\bibnamefont{Dieny}},
  \bibinfo{journal}{Phys. Rev. B} \textbf{\bibinfo{volume}{73}},
  \bibinfo{pages}{064414} (\bibinfo{year}{2006}).

\bibitem[{\citenamefont{Chen et~al.}(2006)\citenamefont{Chen, Rooks, Ruiz, Sun,
  and Kent}}]{chen06}
\bibinfo{author}{\bibfnamefont{W.}~\bibnamefont{Chen}},
  \bibinfo{author}{\bibfnamefont{M.~J.} \bibnamefont{Rooks}},
  \bibinfo{author}{\bibfnamefont{N.}~\bibnamefont{Ruiz}},
  \bibinfo{author}{\bibfnamefont{J.~Z.} \bibnamefont{Sun}}, \bibnamefont{and}
  \bibinfo{author}{\bibfnamefont{A.~D.} \bibnamefont{Kent}},
  \bibinfo{journal}{Phys. Rev. B} \textbf{\bibinfo{volume}{74}},
  \bibinfo{pages}{144408} (\bibinfo{year}{2006}).

\bibitem[{\citenamefont{Tanaka et~al.}(2002)\citenamefont{Tanaka, Shimizu,
  Seyama, Nagasaka, Kondo, Oshima, Eguchi, and Kanai}}]{tanaka02}
\bibinfo{author}{\bibfnamefont{A.}~\bibnamefont{Tanaka}},
  \bibinfo{author}{\bibfnamefont{Y.}~\bibnamefont{Shimizu}},
  \bibinfo{author}{\bibfnamefont{Y.}~\bibnamefont{Seyama}},
  \bibinfo{author}{\bibfnamefont{K.}~\bibnamefont{Nagasaka}},
  \bibinfo{author}{\bibfnamefont{R.}~\bibnamefont{Kondo}},
  \bibinfo{author}{\bibfnamefont{H.}~\bibnamefont{Oshima}},
  \bibinfo{author}{\bibfnamefont{S.}~\bibnamefont{Eguchi}}, \bibnamefont{and}
  \bibinfo{author}{\bibfnamefont{H.}~\bibnamefont{Kanai}},
  \bibinfo{journal}{IEEE Trans. Magn.} \textbf{\bibinfo{volume}{38}},
  \bibinfo{pages}{84} (\bibinfo{year}{2002}).

\bibitem[{\citenamefont{Takagishi et~al.}(2002)\citenamefont{Takagishi, Koi,
  Yoshikawa, Funayama, Iwasaki, and Sahashi}}]{takagishi02}
\bibinfo{author}{\bibfnamefont{M.}~\bibnamefont{Takagishi}},
  \bibinfo{author}{\bibfnamefont{K.}~\bibnamefont{Koi}},
  \bibinfo{author}{\bibfnamefont{M.}~\bibnamefont{Yoshikawa}},
  \bibinfo{author}{\bibfnamefont{T.}~\bibnamefont{Funayama}},
  \bibinfo{author}{\bibfnamefont{H.}~\bibnamefont{Iwasaki}}, \bibnamefont{and}
  \bibinfo{author}{\bibfnamefont{M.}~\bibnamefont{Sahashi}},
  \bibinfo{journal}{IEEE Trans. Magn.} \textbf{\bibinfo{volume}{38}},
  \bibinfo{pages}{2277} (\bibinfo{year}{2002}).

\bibitem[{\citenamefont{Zhu and Zhu}(2004)}]{zhu04}
\bibinfo{author}{\bibfnamefont{J.-G.} \bibnamefont{Zhu}} \bibnamefont{and}
  \bibinfo{author}{\bibfnamefont{X.}~\bibnamefont{Zhu}}, \bibinfo{journal}{IEEE
  Trans. Magn.} \textbf{\bibinfo{volume}{40}}, \bibinfo{pages}{182}
  (\bibinfo{year}{2004}).

\bibitem[{\citenamefont{Grollier et~al.}(2003)\citenamefont{Grollier, Cros,
  Jaffr\'es, Hamzic, George, Faini, Youssef, Gall, and Fert}}]{grollier03}
\bibinfo{author}{\bibfnamefont{J.}~\bibnamefont{Grollier}},
  \bibinfo{author}{\bibfnamefont{V.}~\bibnamefont{Cros}},
  \bibinfo{author}{\bibfnamefont{H.}~\bibnamefont{Jaffr\'es}},
  \bibinfo{author}{\bibfnamefont{A.}~\bibnamefont{Hamzic}},
  \bibinfo{author}{\bibfnamefont{J.~M.} \bibnamefont{George}},
  \bibinfo{author}{\bibfnamefont{G.}~\bibnamefont{Faini}},
  \bibinfo{author}{\bibfnamefont{J.~B.} \bibnamefont{Youssef}},
  \bibinfo{author}{\bibfnamefont{H.} \bibnamefont{LeGall}}, \bibnamefont{and}
  \bibinfo{author}{\bibfnamefont{A.}~\bibnamefont{Fert}},
  \bibinfo{journal}{Phys. Rev. B} \textbf{\bibinfo{volume}{67}},
  \bibinfo{pages}{174402} (\bibinfo{year}{2003}).

\bibitem[{\citenamefont{Morise and Nakamura}(2005)}]{morise05}
\bibinfo{author}{\bibfnamefont{H.}~\bibnamefont{Morise}} \bibnamefont{and}
  \bibinfo{author}{\bibfnamefont{S.}~\bibnamefont{Nakamura}},
  \bibinfo{journal}{Phys. Rev. B} \textbf{\bibinfo{volume}{71}},
  \bibinfo{pages}{014439} (\bibinfo{year}{2005}).

\bibitem[{\citenamefont{Zhang et~al.}(2002)\citenamefont{Zhang, Levy, and
  Fert}}]{zhang02}
\bibinfo{author}{\bibfnamefont{S.}~\bibnamefont{Zhang}},
  \bibinfo{author}{\bibfnamefont{P.~M.} \bibnamefont{Levy}}, \bibnamefont{and}
  \bibinfo{author}{\bibfnamefont{A.}~\bibnamefont{Fert}},
  \bibinfo{journal}{Phys. Rev. Lett.} \textbf{\bibinfo{volume}{88}},
  \bibinfo{pages}{236601} (\bibinfo{year}{2002}).

\bibitem[{\citenamefont{Zhang et~al.}(2004)\citenamefont{Zhang, Levy, Zhang,
  and Antropov}}]{zhang04}
\bibinfo{author}{\bibfnamefont{J.}~\bibnamefont{Zhang}},
  \bibinfo{author}{\bibfnamefont{P.~M.} \bibnamefont{Levy}},
  \bibinfo{author}{\bibfnamefont{S.}~\bibnamefont{Zhang}}, \bibnamefont{and}
  \bibinfo{author}{\bibfnamefont{V.}~\bibnamefont{Antropov}},
  \bibinfo{journal}{Phys. Rev. Lett.} \textbf{\bibinfo{volume}{93}},
  \bibinfo{pages}{256602} (\bibinfo{year}{2004}).

\bibitem[{\citenamefont{Taniguchi et~al.}(2008)\citenamefont{Taniguchi, Yakata,
  Imamura, and Ando}}]{taniguchi08}
\bibinfo{author}{\bibfnamefont{T.}~\bibnamefont{Taniguchi}},
  \bibinfo{author}{\bibfnamefont{S.}~\bibnamefont{Yakata}},
  \bibinfo{author}{\bibfnamefont{H.}~\bibnamefont{Imamura}}, \bibnamefont{and}
  \bibinfo{author}{\bibfnamefont{Y.}~\bibnamefont{Ando}},
  \bibinfo{journal}{Appl. Phys. Express} \textbf{\bibinfo{volume}{1}},
  \bibinfo{pages}{031302} (\bibinfo{year}{2008}).

\bibitem[{\citenamefont{Mizukami
  et~al.}(2002{\natexlab{a}})\citenamefont{Mizukami, Ando, and
  Miyazaki}}]{mizukami02a}
\bibinfo{author}{\bibfnamefont{S.}~\bibnamefont{Mizukami}},
  \bibinfo{author}{\bibfnamefont{Y.}~\bibnamefont{Ando}}, \bibnamefont{and}
  \bibinfo{author}{\bibfnamefont{T.}~\bibnamefont{Miyazaki}},
  \bibinfo{journal}{J. Magn. Magn. Mater.} \textbf{\bibinfo{volume}{239}},
  \bibinfo{pages}{42} (\bibinfo{year}{2002}{\natexlab{a}}).

\bibitem[{\citenamefont{Mizukami
  et~al.}(2002{\natexlab{b}})\citenamefont{Mizukami, Ando, and
  Miyazaki}}]{mizukami02b}
\bibinfo{author}{\bibfnamefont{S.}~\bibnamefont{Mizukami}},
  \bibinfo{author}{\bibfnamefont{Y.}~\bibnamefont{Ando}}, \bibnamefont{and}
  \bibinfo{author}{\bibfnamefont{T.}~\bibnamefont{Miyazaki}},
  \bibinfo{journal}{Phys. Rev. B} \textbf{\bibinfo{volume}{66}},
  \bibinfo{pages}{104413} (\bibinfo{year}{2002}{\natexlab{b}}).

\bibitem[{\citenamefont{Tserkovnyak et~al.}(2002)\citenamefont{Tserkovnyak,
  Brataas, and Bauer}}]{tserkovnyak02}
\bibinfo{author}{\bibfnamefont{Y.}~\bibnamefont{Tserkovnyak}},
  \bibinfo{author}{\bibfnamefont{A.}~\bibnamefont{Brataas}}, \bibnamefont{and}
  \bibinfo{author}{\bibfnamefont{G.~E.~W.} \bibnamefont{Bauer}},
  \bibinfo{journal}{Phys. Rev. B} \textbf{\bibinfo{volume}{66}},
  \bibinfo{pages}{224403} (\bibinfo{year}{2002}).

\bibitem[{\citenamefont{Tserkovnyak et~al.}(2003)\citenamefont{Tserkovnyak,
  Brataas, and Bauer}}]{tserkovnyak03}
\bibinfo{author}{\bibfnamefont{Y.}~\bibnamefont{Tserkovnyak}},
  \bibinfo{author}{\bibfnamefont{A.}~\bibnamefont{Brataas}}, \bibnamefont{and}
  \bibinfo{author}{\bibfnamefont{G.~E.~W.} \bibnamefont{Bauer}},
  \bibinfo{journal}{Phys. Rev. B} \textbf{\bibinfo{volume}{67}},
  \bibinfo{pages}{140404(R)} (\bibinfo{year}{2003}).

\bibitem[{\citenamefont{Brataas et~al.}(2001)\citenamefont{Brataas, Nazarov,
  and Bauer}}]{brataas01}
\bibinfo{author}{\bibfnamefont{A.}~\bibnamefont{Brataas}},
  \bibinfo{author}{\bibfnamefont{Y.~V.} \bibnamefont{Nazarov}},
  \bibnamefont{and} \bibinfo{author}{\bibfnamefont{G.~E.~W.}
  \bibnamefont{Bauer}}, \bibinfo{journal}{Eur. Phys. J. B}
  \textbf{\bibinfo{volume}{22}}, \bibinfo{pages}{99} (\bibinfo{year}{2001}).

\bibitem[{\citenamefont{Zwierzycki et~al.}(2005)\citenamefont{Zwierzycki,
  Tserkovnyak, Kelly, Brataas, and Bauer}}]{zwierzycki05}
\bibinfo{author}{\bibfnamefont{M.}~\bibnamefont{Zwierzycki}},
  \bibinfo{author}{\bibfnamefont{Y.}~\bibnamefont{Tserkovnyak}},
  \bibinfo{author}{\bibfnamefont{P.~J.} \bibnamefont{Kelly}},
  \bibinfo{author}{\bibfnamefont{A.}~\bibnamefont{Brataas}}, \bibnamefont{and}
  \bibinfo{author}{\bibfnamefont{G.~E.~W.} \bibnamefont{Bauer}},
  \bibinfo{journal}{Phys. Rev. B} \textbf{\bibinfo{volume}{71}},
  \bibinfo{pages}{064420} (\bibinfo{year}{2005}).

\bibitem[{\citenamefont{Bauer et~al.}(2003)\citenamefont{Bauer, Tserkovnyak,
  H-Hernando, and Brataas}}]{bauer03}
\bibinfo{author}{\bibfnamefont{G.~E.~W.} \bibnamefont{Bauer}},
  \bibinfo{author}{\bibfnamefont{Y.}~\bibnamefont{Tserkovnyak}},
  \bibinfo{author}{\bibfnamefont{D.}~\bibnamefont{Huertas-Hernando}},
  \bibnamefont{and} \bibinfo{author}{\bibfnamefont{A.}~\bibnamefont{Brataas}},
  \bibinfo{journal}{Phys. Rev. B} \textbf{\bibinfo{volume}{67}},
  \bibinfo{pages}{094421} (\bibinfo{year}{2003}).

\bibitem[{\citenamefont{Slonczewski}(2006)}]{slonczewski06}
\bibinfo{author}{\bibfnamefont{J.~C.} \bibnamefont{Slonczewski}},
  \bibinfo{journal}{Phys. Rev. Lett.} \textbf{\bibinfo{volume}{96}},
  \bibinfo{pages}{019707} (\bibinfo{year}{2006}).

\bibitem[{\citenamefont{Urazhdin et~al.}(2005)\citenamefont{Urazhdin, Loloee,
  and W.~P.~Pratt}}]{urazhdin05}
\bibinfo{author}{\bibfnamefont{S.}~\bibnamefont{Urazhdin}},
  \bibinfo{author}{\bibfnamefont{R.}~\bibnamefont{Loloee}}, \bibnamefont{and}
  \bibinfo{author}{\bibfnamefont{W.}~\bibnamefont{P.~Pratt}},
  \bibinfo{journal}{Phys. Rev. B} \textbf{\bibinfo{volume}{71}},
  \bibinfo{pages}{100401(R)} (\bibinfo{year}{2005}).

\bibitem[{\citenamefont{Guo and Jalil}(2005)}]{guo05}
\bibinfo{author}{\bibfnamefont{J.}~\bibnamefont{Guo}} \bibnamefont{and}
  \bibinfo{author}{\bibfnamefont{M.~B.~A.} \bibnamefont{Jalil}},
  \bibinfo{journal}{Phys. Rev. B} \textbf{\bibinfo{volume}{71}},
  \bibinfo{pages}{224408} (\bibinfo{year}{2005}).

\bibitem[{\citenamefont{Valet and Fert}(1993)}]{valet93}
\bibinfo{author}{\bibfnamefont{T.}~\bibnamefont{Valet}} \bibnamefont{and}
  \bibinfo{author}{\bibfnamefont{A.}~\bibnamefont{Fert}},
  \bibinfo{journal}{Phys.Rev.B} \textbf{\bibinfo{volume}{48}},
  \bibinfo{pages}{7099} (\bibinfo{year}{1993}).

\bibitem[{\citenamefont{Taniguchi and Imamura}(2007)}]{taniguchi07}
\bibinfo{author}{\bibfnamefont{T.}~\bibnamefont{Taniguchi}} \bibnamefont{and}
  \bibinfo{author}{\bibfnamefont{H.}~\bibnamefont{Imamura}},
  \bibinfo{journal}{Phys. Rev. B} \textbf{\bibinfo{volume}{76}},
  \bibinfo{pages}{092402} (\bibinfo{year}{2007}).

\bibitem[{\citenamefont{Bass and Jr}(2007)}]{bass07}
\bibinfo{author}{\bibfnamefont{J.}~\bibnamefont{Bass}} \bibnamefont{and}
  \bibinfo{author}{\bibfnamefont{W.~P.~P.} \bibnamefont{Jr}},
  \bibinfo{journal}{J. Phys.: Condens. Matter} \textbf{\bibinfo{volume}{19}},
  \bibinfo{pages}{183201} (\bibinfo{year}{2007}).

\bibitem[{\citenamefont{Beaujour et~al.}(2006)\citenamefont{Beaujour, Chen,
  Kent, and Sun}}]{beaujour06}
\bibinfo{author}{\bibfnamefont{J.-M.~L.} \bibnamefont{Beaujour}},
  \bibinfo{author}{\bibfnamefont{W.}~\bibnamefont{Chen}},
  \bibinfo{author}{\bibfnamefont{A.~D.} \bibnamefont{Kent}}, \bibnamefont{and}
  \bibinfo{author}{\bibfnamefont{J.~Z.} \bibnamefont{Sun}},
  \bibinfo{journal}{J. Appl. Phys.} \textbf{\bibinfo{volume}{99}},
  \bibinfo{pages}{08N503} (\bibinfo{year}{2006}).

\bibitem[{\citenamefont{Xia et~al.}(2002)\citenamefont{Xia, Kelly, Bauer,
  Brataas, and Turek}}]{xia02}
\bibinfo{author}{\bibfnamefont{K.}~\bibnamefont{Xia}},
  \bibinfo{author}{\bibfnamefont{P.~J.} \bibnamefont{Kelly}},
  \bibinfo{author}{\bibfnamefont{G.~E.~W.} \bibnamefont{Bauer}},
  \bibinfo{author}{\bibfnamefont{A.}~\bibnamefont{Brataas}}, \bibnamefont{and}
  \bibinfo{author}{\bibfnamefont{I.}~\bibnamefont{Turek}},
  \bibinfo{journal}{Phys. Rev. B} \textbf{\bibinfo{volume}{65}},
  \bibinfo{pages}{220401(R)} (\bibinfo{year}{2002}).

\bibitem[{\citenamefont{Moriyama et~al.}(2008)\citenamefont{Moriyama, Cao, Fan,
  Xuan, Nikoli\'c, Tserkovnyak, Kolodzey, and Xiao}}]{moriyama08}
\bibinfo{author}{\bibfnamefont{T.}~\bibnamefont{Moriyama}},
  \bibinfo{author}{\bibfnamefont{R.}~\bibnamefont{Cao}},
  \bibinfo{author}{\bibfnamefont{X.}~\bibnamefont{Fan}},
  \bibinfo{author}{\bibfnamefont{G.}~\bibnamefont{Xuan}},
  \bibinfo{author}{\bibfnamefont{B.~K.} \bibnamefont{Nikoli\'c}},
  \bibinfo{author}{\bibfnamefont{Y.}~\bibnamefont{Tserkovnyak}},
  \bibinfo{author}{\bibfnamefont{J.}~\bibnamefont{Kolodzey}}, \bibnamefont{and}
  \bibinfo{author}{\bibfnamefont{J.~Q.} \bibnamefont{Xiao}},
  \bibinfo{journal}{Phys. Rev. Lett.} \textbf{\bibinfo{volume}{100}},
  \bibinfo{pages}{067602} (\bibinfo{year}{2008}).

\bibitem[{\citenamefont{Wang et~al.}(2006)\citenamefont{Wang, Bauer, van Wees,
  Brataas, and Tserkovnyak}}]{wang06}
\bibinfo{author}{\bibfnamefont{X.}~\bibnamefont{Wang}},
  \bibinfo{author}{\bibfnamefont{G.~E.~W.} \bibnamefont{Bauer}},
  \bibinfo{author}{\bibfnamefont{B.~J.} \bibnamefont{van Wees}},
  \bibinfo{author}{\bibfnamefont{A.}~\bibnamefont{Brataas}}, \bibnamefont{and}
  \bibinfo{author}{\bibfnamefont{Y.}~\bibnamefont{Tserkovnyak}},
  \bibinfo{journal}{Phys. Rev. Lett.} \textbf{\bibinfo{volume}{97}},
  \bibinfo{pages}{216602} (\bibinfo{year}{2006}).

\bibitem[{\citenamefont{Tserkovnyak et~al.}(2008)\citenamefont{Tserkovnyak,
  Moriyama, and Xiao}}]{tserkovnyak08}
\bibinfo{author}{\bibfnamefont{Y.}~\bibnamefont{Tserkovnyak}},
  \bibinfo{author}{\bibfnamefont{T.}~\bibnamefont{Moriyama}}, \bibnamefont{and}
  \bibinfo{author}{\bibfnamefont{J.~Q.} \bibnamefont{Xiao}},
  \bibinfo{journal}{Phys. Rev. B} \textbf{\bibinfo{volume}{78}},
  \bibinfo{pages}{020401(R)} (\bibinfo{year}{2008}).

\end{thebibliography}

\end{document}